\documentclass[entropy,article,submit,moreauthors,pdftex,10pt,a4paper]{mdpi} 
\usepackage{bm,graphicx,graphics,amsmath,amssymb,epsfig,color}
\preto{\abstractkeywords}{\nolinenumbers}

\def \be {\begin{equation}}
\def \ee {\end{equation}}
\def \beA {\begin{eqnarray}}
\def \eeA {\end{eqnarray}}

\def \average#1{\left\langle #1 \right\rangle}

\usepackage[cmyk,dvipsnames]{xcolor}
\definecolor{pacificb}{HTML}{1CA9C9}

\firstpage{1} 
\articlenumber{x}
\doinum{10.3390/------}
\pubvolume{xx}
\pubyear{2017}
\copyrightyear{2017}
\externaleditor{Academic Editors: C. Meija-Monasterio and L. Rondoni}
\history{Received: date; Accepted: date; Published: date}

\pdfoutput=1

\Title{Dephasing-Assisted Macrospin Transport}

\Author{S. Iubini $^{1,2,*}$, S. Borlenghi $^{3}$, A. Delin$^{3,4}$, S. Lepri $^{1,5}$ and F. Piazza$^{6,7}$}

\AuthorNames{Stefano Iubini, Simone Borlenghi, Anna Delin, Stefano Lepri, Francesco Piazza}

\address{%
$^{1}$ \quad Consiglio Nazionale delle Ricerche, 
Istituto dei Sistemi Complessi, via Madonna del Piano 10, I-50019 Sesto Fiorentino, Italy; {stefano.lepri@isc.cnr.it (S.L.)}\\
$^{2}$\quad Dipartimento di Fisica e Astronomia, Universit\`a di Padova, via F. Marzolo 8 I-35131, Padova, Italy \\
$^{3}$ \quad Department of Applied Physics, School of Engineering Sciences, KTH Royal Institute of Technology, Electrum 229, SE-16440 Kista, Sweden; simonebg@kth.se (S.B.); annadel@kth.se (A.D.)\\
$^{4}$ \quad Department of Physics and Astronomy, Materials Theory Division, Uppsala University, Box 516, SE-75120 Uppsala, Sweden\\
$^{5}$ \quad Istituto Nazionale di Fisica Nucleare, Sezione di Firenze, 
via G. Sansone 1 I-50019, Sesto Fiorentino, Italy\\
$^{6}$ \quad Centre de Biophysique Mol\'eculaire, (CBM), CNRS-UPR 4301, Rue C. Sadron, 45071 Orl\'eans, France; francesco.piazza@cnrs-orleans.fr (F.P.)\\
$^{7}$ \quad Universit\'e d'Orl\'eans, Ch\^ateau de la Source, 45071 Orl\'eans Cedex, France\\
}

\corres{Correspondence: stefano.iubini@fi.isc.cnr.it}

\abstract{
Transport phenomena are ubiquitous in physics, and it is generally understood that the environmental disorder and noise deteriorates the transfer of excitations. There are, however,
cases in which transport can be enhanced by fluctuations.
In the present work, we show, by means of micromagnetics simulations, that transport efficiency in a chain of classical macrospins can be greatly increased by an optimal level of dephasing noise. We also demonstrate the same effect in a simplified model,
the dissipative Discrete Nonlinear Schr\"odinger equation, subject to phase noise.
Our results point towards the realization of a large class of magnonics and spintronics devices, where disorder and noise can be used to enhance spin-dependent transport efficiency.
} 



\begin{document}



\section{Introduction}

Dissipation and fluctuations due to interaction between a system and the environment
play a crucial role in transport and relaxation processes. The understanding of 
the resulting non-equilibrium states is one of the main aims of contemporary
statistical mechanics \cite{Livi2017}. In addition, small open systems display many special properties 
with respect to their macroscopic counterparts due to finite-size fluctuations, reduced 
dimensionality, boundary effects, disorder, etc.~\cite{Puglisi2018,Lepri2016}. Among countless examples of applications, we may mention heat transfer in nanoscale systems and nanophononics \cite{Volz2016}.
In addition, in the last three decades, magnonics and spintronics emerged as new research fields that aim at using the spin degree of freedom in electronics devices \cite{wolf01,kruglyak10} in the form of spin waves and spin dependent electrical currents.
This field has grown enormously in recent years, since it offers promising opportunities for reliable ultrafast nano-size electronics that can be controlled by a variety of means, including electrical currents, magnetic fields, and temperature gradients. 

Spin-caloritronics \cite{uchida08,uchida10,bauer12} concerns precisely the coupled spin/heat transport in systems with non-uniform temperature. The wide interest in 
these types of setups is due to their potential for energy efficient electronics, where heat flows can be used to control information \cite{borlenghi14a,borlenghi18a}.
However, in magnonic devices, especially arrays of nano disks and spin-transfer nano oscillators \cite{slavin09}, the sample-to-sample variability and the noise from the environment are deleterious for the synchronization and transport performance.

On general grounds, there are many situations in which stochastic fluctuations may play a constructive role in enhancing the response of nonlinear systems to an external coherent 
driving. Known examples are the enhancement of the decay time from a meta-stable state 
(noise-enhanced stability \cite{Mantegna1996}), the synchronization with a weak periodic input signal (stochastic resonance \cite{Gammaitoni1998}), the regularizaton of the response at a given optimal noise intensity like coherence resonance \cite{Pikovsky1997}) and resonant 
activation \cite{Doering1992}.

{ 
In the context of quantum transport, the interplay between coherence, structural disorder and noise is a prominent topic of
 investigation}~\cite{schwarzer72,haken73,logan87,logan87b}.
A decade ago, it was shown that transport in quantum dissipative networks can be enhanced by the presence of pure dephasing noise \cite{plenio08,rebentrost09}.
This counter-intuitive phenomenon occurs in disordered systems,
where different local frequencies suppress coherent transport, and a certain level of dephasing noise can broaden the resonance lines of neighboring sites, allowing for the population transfer.
Moreover, the effect appears as a general feature of quantum systems, which can be described using the language of (Markovian or not) quantum master equations. 
These discoveries fostered an intense research activity, aimed at understanding and exploiting dephasing in a variety of systems, including quantum dots \cite{pulido14}, optical fibers \cite{viciani15}, 
photosynthetic reactions \cite{caruso09,chin10,iubini15} and other biological systems \cite{lambert12}.
Recently, noise-assisted transport has been studied experimentally in a chain of trapped atomic ions \cite{roos19}. 

In this paper, we show that dephasing-assisted spin transport occurs in a chain of \emph{classical} coupled (macro)spins governed by the Landau-Lifshitz-Gilbert (LLG) equation. 
The system is described in Section \ref{sec:model}.
In this context the effect emerges from the competition between phase coherence and phase noise and does not require intrinsically quantum-mechanical effects. 
The effect will be first demonstrated in Section \ref{sec:micro} by micromagnetic simulations of the 
full LLG equation. To get further insights, we give in Section \ref{sec:dnls} 
a reduced effective description 
through the non-equilibrium 
discrete nonlinear Schr\"odinger equation (DNLS) \cite{iubini12,iubini13,Mendl2015,Kulkarni2015} that constitutes an approximation of the LLG equation for small amplitude deviations of the spins from their equilibrium position \cite{slavin09}.

\section{The System: Microscopic and Macroscopic Descriptions}
\label{sec:model}

\begin{figure}
\begin{center}
\includegraphics[width=10cm]{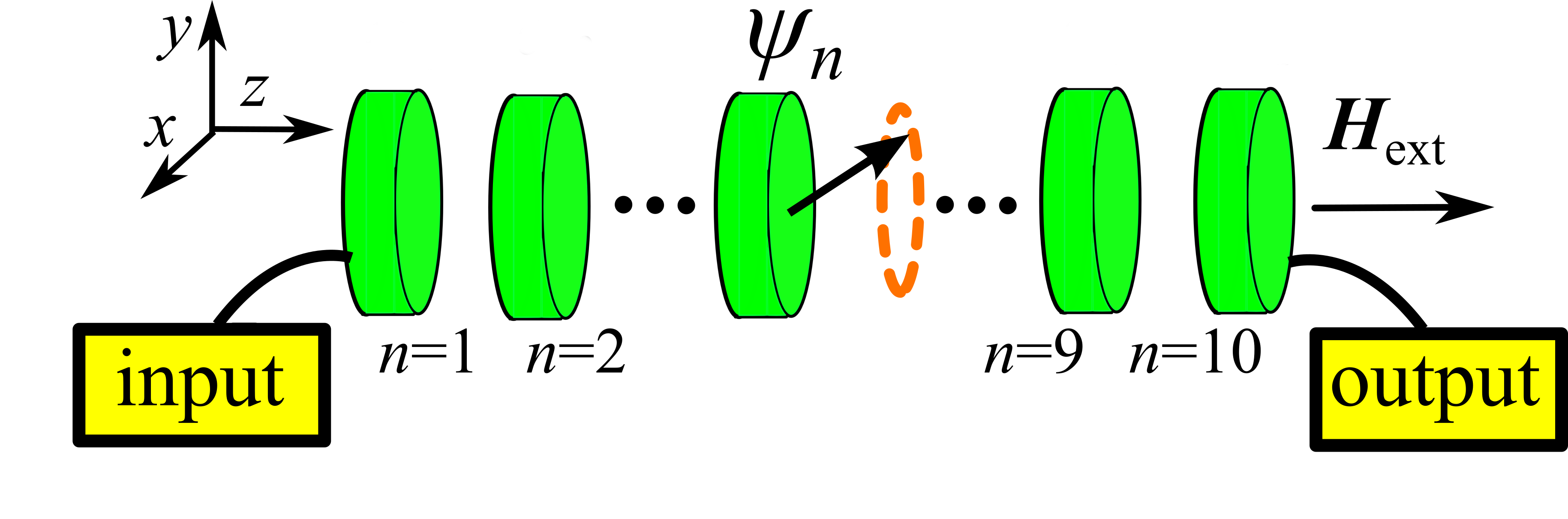}
\end{center}
\caption{Sketch of the system, consisting of 10 Py nano-disks coupled through the magneto-dipolar interaction. The magnetization in the first disk (input) is tilted away from equilibrium. The transport efficiency
is the integrated Spin Wave (SW) power in the last disk (output). 
}
\label{fig:figure1p}
\end{figure}

The system studied here consists of a chain of $N=10$ Permalloy (Py) nano disks coupled through dipolar magnetic interaction, shown in Figure \ref{fig:figure1p}. 
Hereafter, we briefly describe the dynamics of the chain; we refer to References~\cite{gilbert55,slavin09,borlenghi15b} for a thorough discussion.
The dynamics of the magnetization $\bm{M}(\bm{r},t)$ depends on the position $\bm{r}$ inside each of the $n=1,...,N$ disks and is described by the LLG equation of motion:
 
\be\label{eq:llg}
\dot{\bm{M}}=\gamma\bm{M}\times{\bm{H}}_{\rm{eff}}+\frac{\alpha}{M_s}\bm{M}\times\dot{\bm{M}}.
\ee

The first term in Equation~({\ref{eq:llg}}), proportional to the gyromagnetic ratio $\gamma$, describes the precession of the magnetization vector $\bm{M}(\bm{r},t)$ around the effective field 
$\bm{H}_{\rm{eff}}$, while the second term accounts for energy dissipation at a rate proportional to the phenomenological Gilbert damping parameter $\alpha$. The saturation magnetization $M_s$ is the norm of the magnetization, conserved during the dynamics, which depends on the material properties and sample geometry.

The effective field 
\be\label{eq:fields}
\bm{H}_{\rm{eff}}=\bm{H}_{\rm{ext}}+\bm{H}_{\rm{exc}}+\bm{H}_{\rm{dip}}+\bm{H}_{\rm{th}}
\ee
is the sum of four contributions. First comes the 
applied field $\bm{H}_{\rm{ext}}=H\hat{\bm{z}}$, which defines the precession axis of the magnetization $\bm{\hat{z}}$. Then, the exchange field $\bm{H}_{\rm{exc}}$ (proportional to the exchange stiffness $A$) is the short-range interaction
that accounts for the coherent precession of the magnetization inside each disk.
%
The third term is the dipolar field $\bm{H}_{\rm{dip}}$, which contains contributions from volume and surface charges. The dipolar field acts as a demagnetizing field in each disk and is responsible for the coupling between the disks and the nonlinearity of the dynamics \cite{slavin09}.
Thermal fluctuations in each disk are modeled by the stochastic field
\be\label{eq:thermal_field}
\bm{H}_{\rm{th}}=\sqrt{DT}[h_x^{\rm{th}}({\bm{r}},t),h_y^{\rm{th}}({\bm{r}},t),h_z^{\rm{th}}({\bm{r}},t)].
\ee 

Here, $T$ is the temperature of the thermal bath and
\be\label{eq:diffusion}
D=\frac{2\alpha k_B}{\gamma\mu_0 VM_s},
\ee
where $k_B$ is the Boltzmann constant, $\mu_0$ the vacuum magnetic permeability, and $V$ the elementary volume containing the magnetization vector $\bm{M}$. In finite-element micromagnetics simulations, the latter corresponds to the volume of the mesh elements.
Each component of the thermal field is modeled as a Gaussian random variable with zero average and correlation $\average{h_i^{\rm{th}}(\bm{r},t),h_j^{\rm{th}}(\bm{r},t^\prime)}=\delta(\bm{r}-\bm{r^\prime})\delta(t-t^\prime)\delta_{ij}$ with $i,j=x,y,z$. 

In this paper, we consider a chain of Permalloy nanodisks with thickness $t = 3$ nm, radius $R = 20$~nm, and an interlayer distance $d = 3$~nm. The applied field $\bm{H}_{\rm{ext}} = 1$ T defines the precession axis of the 
magnetization along the $\bm{\hat{z}}$ direction. The exchange stiffness $A = 10^{-11}$ J/m corresponds to that of Permalloy, while the other micromagnetics parameters are
$M_s=0.94$T/$\mu_0$, $\alpha=8\times 10^{-3}$, and $\gamma= 1.873 \times 10^{11}$ rad s$^{-1}$ T$^{-1}$. Those parameters are taken from Reference~\cite{naletov11}.
The evolution of the disk chain was performed at the microscopic level 
by solving Equation (\ref{eq:llg}) with the NMAG software~\cite{fischbacher07}, using a tetrahedral finite element mesh with maximum size of 3 nm, of the order of the Permalloy exchange length.

In nano disks with radius of a few hundreds of nm, the magnetization precession typically exhibits several precession normal modes \cite{naletov11,lupo15}.
For significantly smaller nanodisks, like in our case, one can verify that the only mode that is
effectively active is the uniform one (i.e., a uniform, phase coherent magnetization precession over the disk volume) \cite{lupo15}.
In this regime the system can be reduced to an ensemble of coupled macrospins defined as $\bm{M}^n(t)=\frac{1}{V_n}\int_{V_n}\bm{M}(\bm{r}_n,t){\mbox{d}}^3r_n$. 
Note that here $V_n$ is the volume of each disk and
 not the volume element that enters the diffusion constant in Equation (\ref{eq:diffusion}).

 It is thus convenient to introduce the complex Spin Wave (SW) amplitude \cite{slavin09}: 
\be
\label{eq:psi}
\psi_n=\frac{M_x^n+iM_y^n}{\sqrt{2M_s(M_s+M_z^n)}}.
\ee
By writing Equation~(\ref{eq:psi}) as $\psi_n=\sqrt{p_n(t)}\mbox{e}^{i\phi_n(t)}$, one can see that the phase
$\phi_n(t)$ describes the precession of $\bm{M}^n$ in the $x$-$y$ plane, while $p_n=|\psi_n|^2$ is the local SW power. From these definitions, it follows that the
equilibrium solution $\bm{M}^n=M_s \bm{\hat{z}}$ corresponds to a vanishing SW field with $p_n=0$.
More in general, the macroscopic spin dynamics of the disk chain can be mapped to the evolution of a system of damped oscillators described by complex variables \cite{borlenghi15b}, as discussed in Section \ref{sec:dnls}.

Before presenting the main results, it is useful to understand the nature of the 
collective excitations of the disk array. A way to evaluate empirically the 
spin-wave spectrum is to first determine the main frequency peaks $\omega_n$ appearing 
in the Fourier transform of the total magnetization $\sum_n \bm{M}^n$, as done in Reference~\cite{borlenghi15b}.
From micromagnetics simulations with the parameters given above, we found that
there are five main components at frequencies $\omega_1=18.0$, $\omega_2=19.7$, $\omega_3=20.5$,
$\omega_4=21.8$, and $\omega_5=24.0$ GHz.
Since each disk behaves as a magnetic dipole, aligning the disks in a chain gives a structure where the intensity of the dipolar field, which controls the frequencies, is symmetric around the centre of the chain, yielding a double degeneracy of the 
spectrum.

\begin{figure}[H]
\begin{center}
\includegraphics[width=10cm]{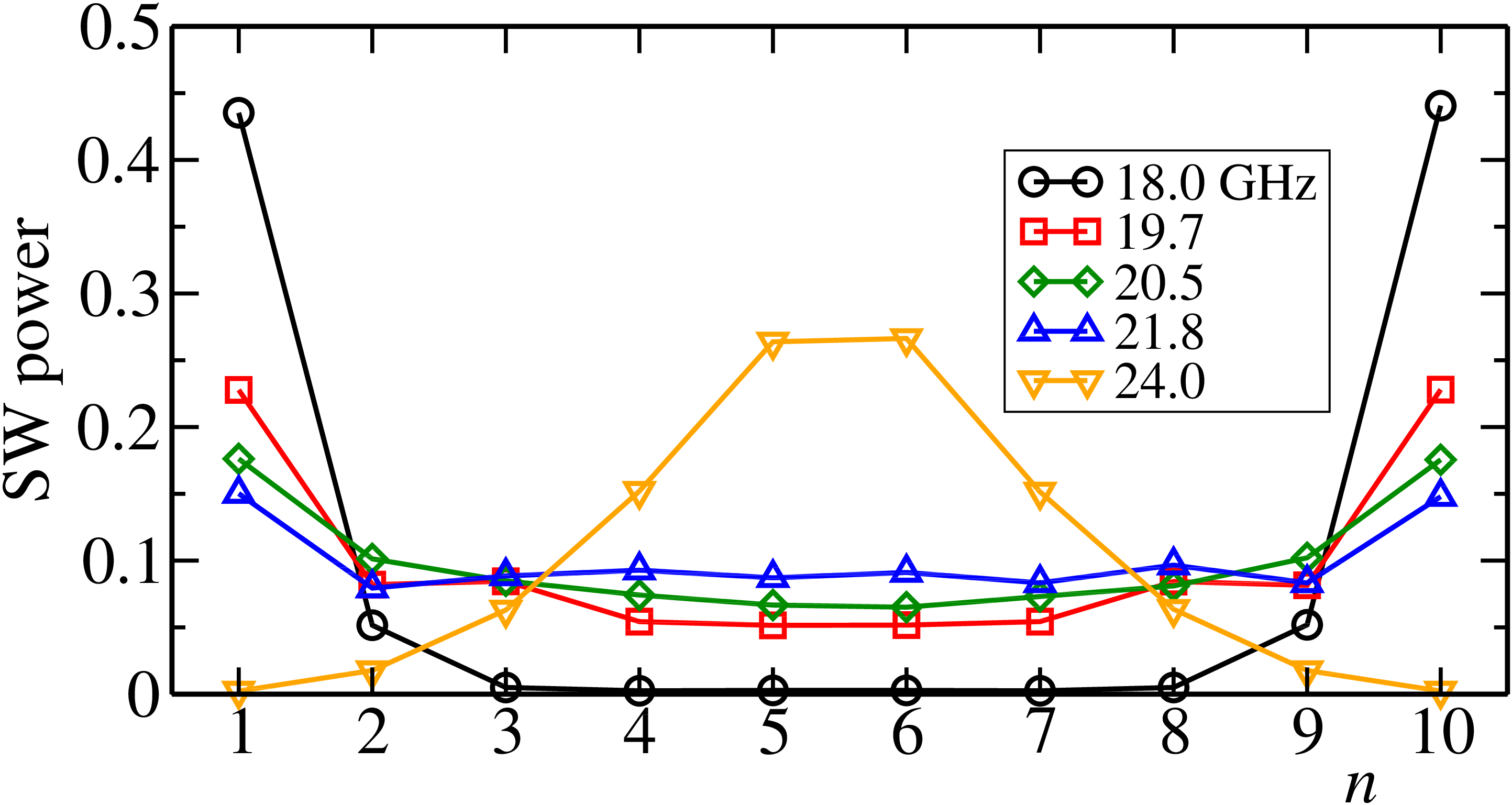}
\end{center}
\caption{Time-average of the SW powers of each collective mode, obtained by exciting the dynamics with a uniform time-dependent magnetic field with the frequencies of the modes until the system reaches a steady state. Simulations are performed at zero temperature. The total SW power of each profile is normalized to one for better comparison. }
\label{fig:figure2p}
\end{figure}

To determine the spatial structure of the spin-wave modes in an effective way, we
apply a uniform time-dependent magnetic field oscillating at the frequencies of the
modes and measure the average profile of the spin-wave powers $p_n$ 
in the steady state. The simulations were performed at zero temperature.
The resulting profiles are displayed in Figure \ref{fig:figure2p}.
The profiles are symmetric with respect to the chain center. 
Moreover the mode with the highest-frequency $\omega_5=24.0$ GHz differs from
the others, being localized on the two central disks.
Although the disk chain is not an intrinsically disordered system, the quite well localized 
nature of its collective modes indicates that effective SW transport is limited to a restricted 
dynamical regime where the macrospins perform coherent nonlocal oscillations. 
On the other hand, spatially localized excitations are quickly damped by the Gilbert dissipative term in Equation~(\ref{eq:llg}).

\section{Micromagnetics Simulations with Dephasing Noise}
\label{sec:micro}

Here, we are interested on the effect of pure dephasing noise on the disks dynamics. 
In principle, this could be physically realized by introducing a suitable 
random modulation of the external applied field, varying on much faster time-scales
with respect to the typical precession period of the macrospins.
The strength of the randomness should be the main accessible control parameter.
To implement dephasing in the model,
at each time step $\tau$ of the numerical solution of the LLG equation, we transform the magnetization in 
each disk as $\bm{M}^n\rightarrow R_z(\Theta_n)\bm{M}^n$, where $R_z$ is a 
rotation matrix by a random angle $\Theta_n$ 
 around the $z$ axis ($R_z(0)$ being 
the identity matrix). The angles 
$\Theta_n$ 
are independent, identically distributed Gaussian variables with
zero average and a standard deviation equal to $\theta c\sqrt{\tau}$,
where $\theta$ is an non-dimensional parameter which
controls the strength of the dephasing noise and $c=1\,deg\,ps^{-1/2}$. 
By construction, such a process 
exactly conserves the spin powers.

Note that, at variance with previous studies \cite{borlenghi15b,borlenghi16a}, here we do not consider off-equilibrium steady states, but we study transport in a transient regime. 
In particular, we start from a condition where only the magnetization of the first disk
is not aligned with the $\bm{\hat{z}}$ axis, and we monitor the evolution of the system as the 
energy pulse propagates through the chain, until $\bm{M}$ is aligned with the $\bm{z}$ axis in all the disks. More precisely, the initial condition is generated by tilting the magnetization of the first disk in the $\bm{\hat{x}}$ direction by a given angle,
leaving the other magnetizations aligned with $\bm{\hat{z}}$ and monitoring the time evolution for
about 10 ns with a time step of 1 ps. Unless otherwise stated, all the simulations were performed 
setting the bath temperature $T=0$
in Equation~(\ref{eq:thermal_field}) and changing the noise strength $\theta$.

\begin{figure}[H]
\begin{center}
\includegraphics[width=10cm]{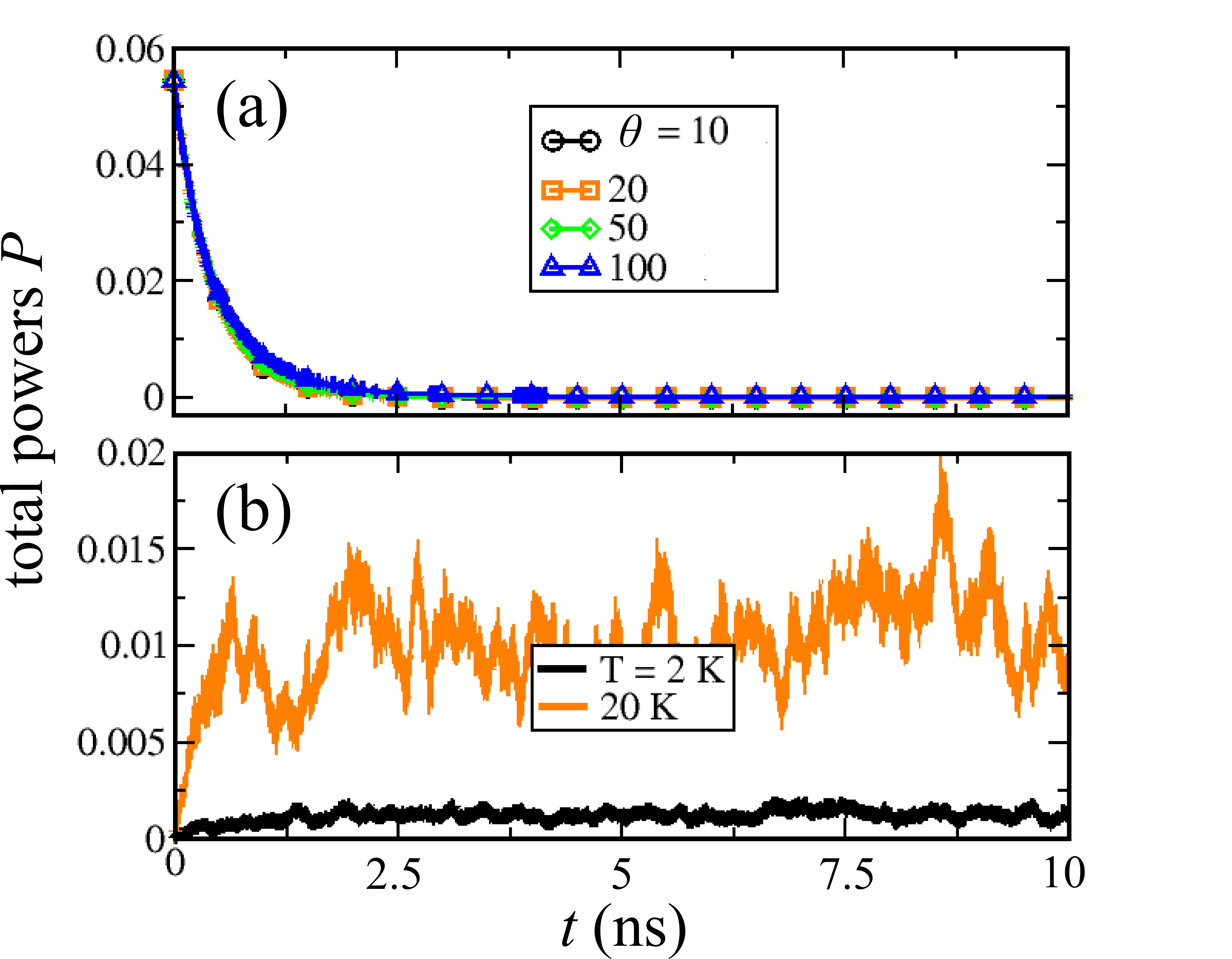}
\end{center}
\caption{Total SW power $P$ vs time for different values of the dephasing 
noise amplitude $\theta$ (\textbf{a}) and of the bath temperature $T$ (\textbf{b}). One can see that in the first case $P$ drops to zero and the magnetization aligns
with the $z$ axis, since the dephasing conserves the total power. In the second case, the bath temperature excites the dynamics and the system thermalizes with $P$ increasing
with the bath temperature $T$.}
\label{fig:figure3p}
\end{figure}

To understand better the nature of the dephasing process, 
we first compare the dynamics of the system in contact with the thermal bath 
only, as given by Equation~(\ref{eq:thermal_field}) ($\theta=0$, $T\neq 0$), 
with the one in presence of a pure dephasing noise ($\theta\neq 0$, $T = 0$). 
Figure \ref{fig:figure3p} 
compares the time evolution of the total SW power $P=\sum_np_n$ in the two cases. 
From Figure \ref{fig:figure3p}a, one can see that the strength of the dephasing process
hardly changes the relaxation to the equilibrium state $P=0$, which is 
solely determined by intrinsic dissipation. On the other hand, 
thermal fluctuations affect the dynamics in a completely different
fashion by making the precession amplitudes fluctuate steadily as expected. 

\begin{figure}[H]
\begin{center}
\includegraphics[width=12cm]{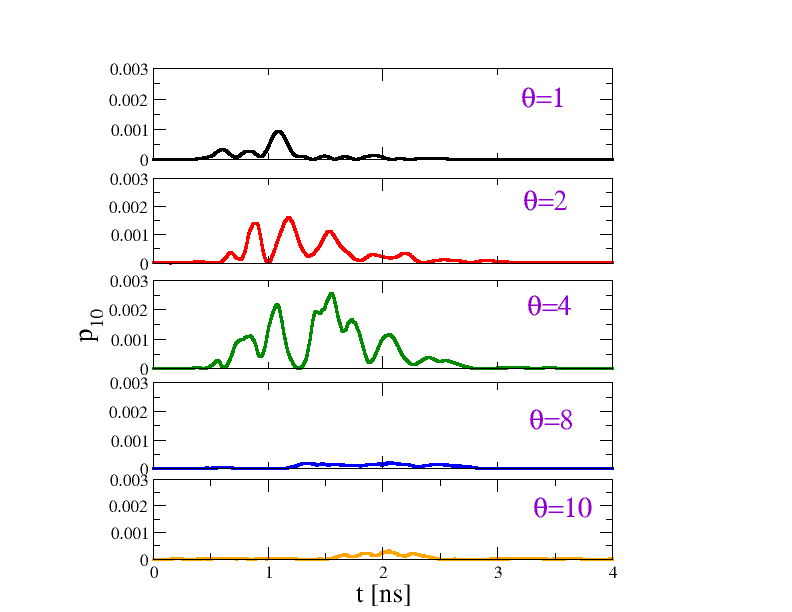}
\end{center}
\caption{Effect of the dephasing noise on the dynamics of the macrospin chain: 
time evolution of the local SW power $p_{10}$ of the last disk for different values
of $\theta$. Transmitted power is maximized for an optimal value around 
$\theta\approx 4$. Simulation parameters as given in the text. 
}
\label{fig:figure4p}
\end{figure}

\begin{figure}[H]
\begin{center}
\includegraphics[width=10cm]{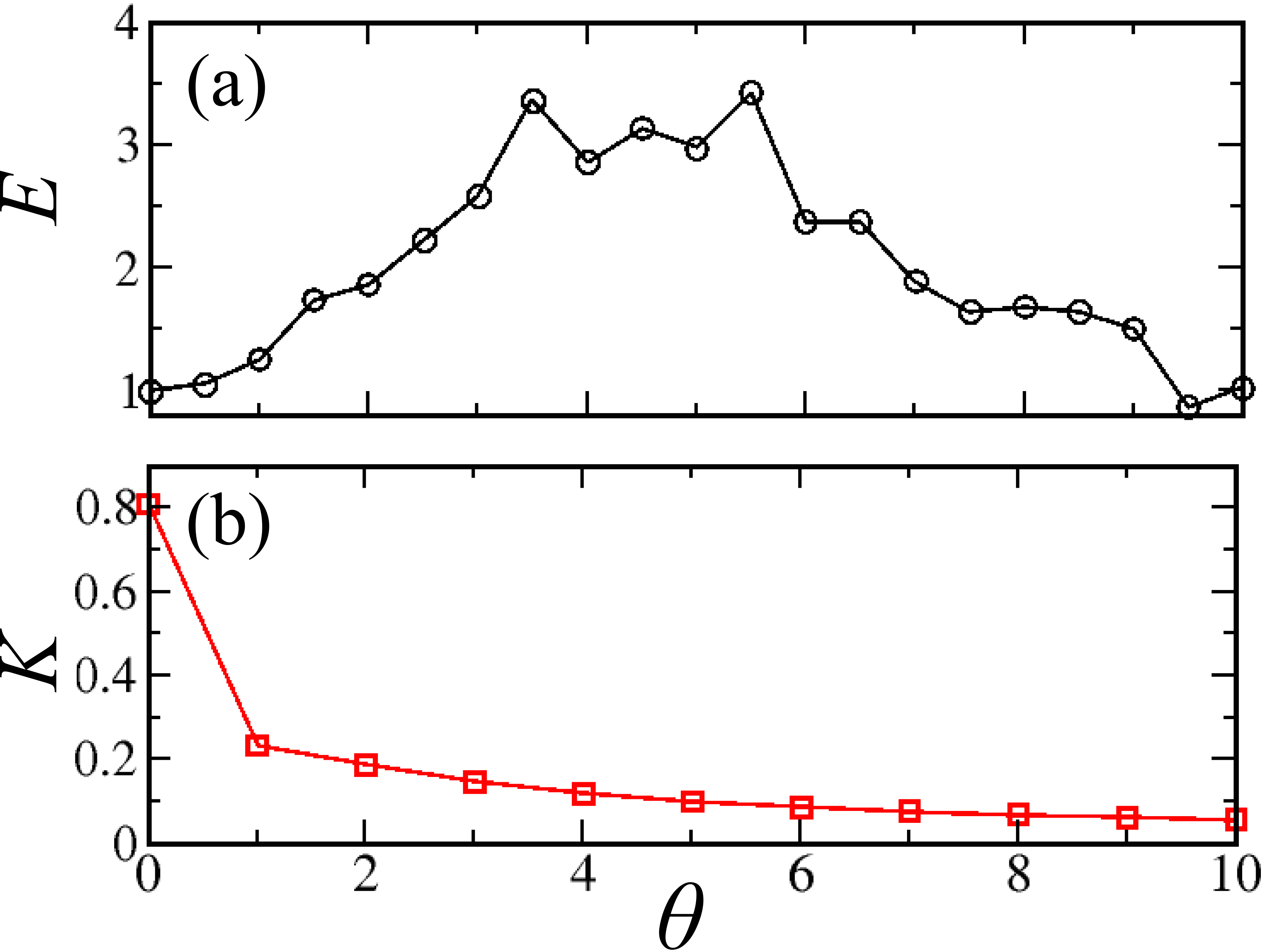}
\end{center}
\caption{(\textbf{a}) Efficiency $E$ and (\textbf{b}) Kuramoto parameter $K$ versus $\theta$. $E$ increases of a factor 3 until $\theta=6$ and then decreases again, showing that transport can be effectively promoted
by dephasing. On the other hand, $K$ decreases monotonically with $\theta$. Thus, in the present case, transport is not related to phase synchronization.
}
\label{fig:figure5p}
\end{figure}

Let us now turn to the effect of the dephasing on transport.
Figure {\ref{fig:figure4p}} shows the time evolution of the SW power of the last disk as a function of time for increasing $\theta$. One can see clearly that the pulse increases up to $\theta \approx 4$ and then decreases.
To quantify transport efficiency, we compare the total power that flows through the 
last disk, and we compute
\be\label{eq:efficiency}
E=\frac{\int_0^\infty p_{10}(t,\theta)dt}{\int_0^\infty p_{10}(t,\theta=0)dt}.
\ee
A value $E>1$ is indicative of an enhancement or energy 
transfer induced by the dephasing process with respect to the noise-free case.
In Figure \ref{fig:figure5p}a, we report the efficiency $E$, defined in Equation (\ref{eq:efficiency})
as a function of the parameter $\theta$; one can see that the $E$ reaches the maximum between $\theta=4$ and 6, and then decreases. In other words, there is an optimal range of 
$\theta$ where transport is enhanced with respect to the noiseless case.

In previous studies~\cite{borlenghi14a,borlenghi16a}, it has been shown that the current increase at increasing phase synchronization
 and transport is due to phase coherence.
However, in the present case, transport is mostly incoherent. This is seen in 
Figure~\ref{fig:figure5p}b, which shows the Kuramoto synchronization parameter~\cite{kuramoto1975self,kuramotochemical}, defined as 
$K=\langle\overline{\frac{1}{N}|\sum_{n=1}^N e^{i\phi_n(t)}|}\rangle$, 
where $\langle \cdot \rangle$ and $\overline{\,\cdot\,}$ denote, respectively, ensemble and time averages and $\phi_n(t)$ is the time-dependent phase of the $n^{th}$ oscillator. 
At variance with the efficiency, $K$ does not display any maximum but rather decreases monotonically as dephasing destroys the phase coherence among magnetization vectors
of the the disks. We remark that, as already illustrated from Figure \ref{fig:figure3p}, the increase of transport efficiency is not trivially 
due to the fact that more SW power is injected into the system, since dephasing noise conserves the total power. 
This indicates
that a different mechanism needs to be invoked to explain the increase of the efficiency with noise.

In order to get some insight on such mechanisms, we consider the spectral content of the magnetization in the different regimes.
A convenient way to analyze non-stationary signals is to use wavelet analysis in the time-frequency domain. 
This method allows to detect transient frequency components appearing at specific times and lasting for finite lapses of time. 
In this work, we computed the Gabor transform~\cite{Gabor:1946aa} of 
the complex magnetization of a given disk $n$, namely
\begin{equation}
\label{e:gabor}
G_n(\omega,t) = \int_{-\infty}^{+\infty} e^{-(t-\tau)^2/a} e^{-i\omega \tau } \psi_n(\tau ) \, d\tau .
\end{equation}

\begin{figure}[H]
\begin{center}
\includegraphics[width=0.8\textwidth]{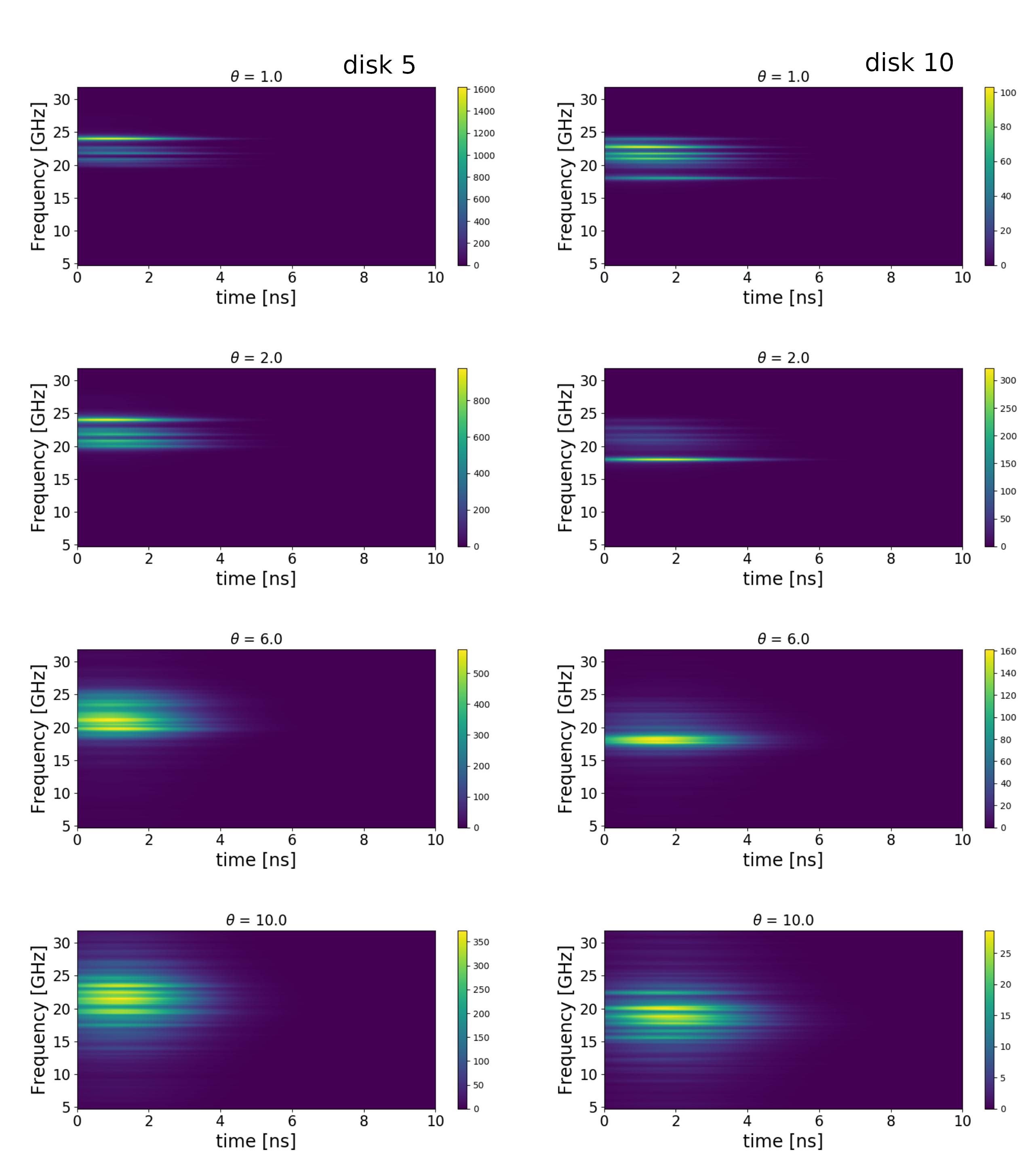}
\end{center}
\caption{Wavelet analysis of the complex spin amplitudes $\psi_n$ on the central disk ($n=5$, left panels) and
on the last disk ($n=10$, right panels) for different values of $\theta$. Each plot shows the density map of the 
average square modulus $\langle |G_n(\omega,t)|^2\rangle $ of the Gabor transform (\ref{e:gabor}) averaged over a sample 
of 32 independent realizations of the dyanamics.
The parameter $a$ has been set equal to $7.5$ ns$^2$, optimized so as to maximize the resolution 
in both the time and frequency domains. Notice the difference in the density scales.
 }
\label{fig:figure6p}
\end{figure}

\noindent It is interesting to compare the behavior of the wavelet signal at two different positions of the chain, namely
the bulk and the output region. 
Figure~\ref{fig:figure6p} shows the average density maps $\langle |G_n(\omega,t)|^2 \rangle$ for different values
of the noise strength $\theta$ in the interval $[1, 10]$ for disks $n=5$ and $n=10$.
In the whole temperature range, the wavelet signal lasts approximately for $5\,ns$ before being damped, which is 
the same typical damping time observed in Figure~\ref{fig:figure4p} for the local power at the chain end. 
For small $\theta$, we can distinguish a dominant mode excited at $\omega=\omega_5=24$ GHz localized on the central site $n=5$
and a mode at $\omega_1=18$ GHz localized on $n=10$. This structure reflects the pattern of collective modes shown in 
Figure~\ref{fig:figure2p}.
At larger $\theta$ values, the number of transmitting modes in the frequency range around $\omega = 20$ GHz increases. A further 
increment of the noise strength
broadens and weakens the resonance lines, yielding the efficiency reduction described above.

\begin{figure}[H]
\begin{center}
\includegraphics[width=10cm]{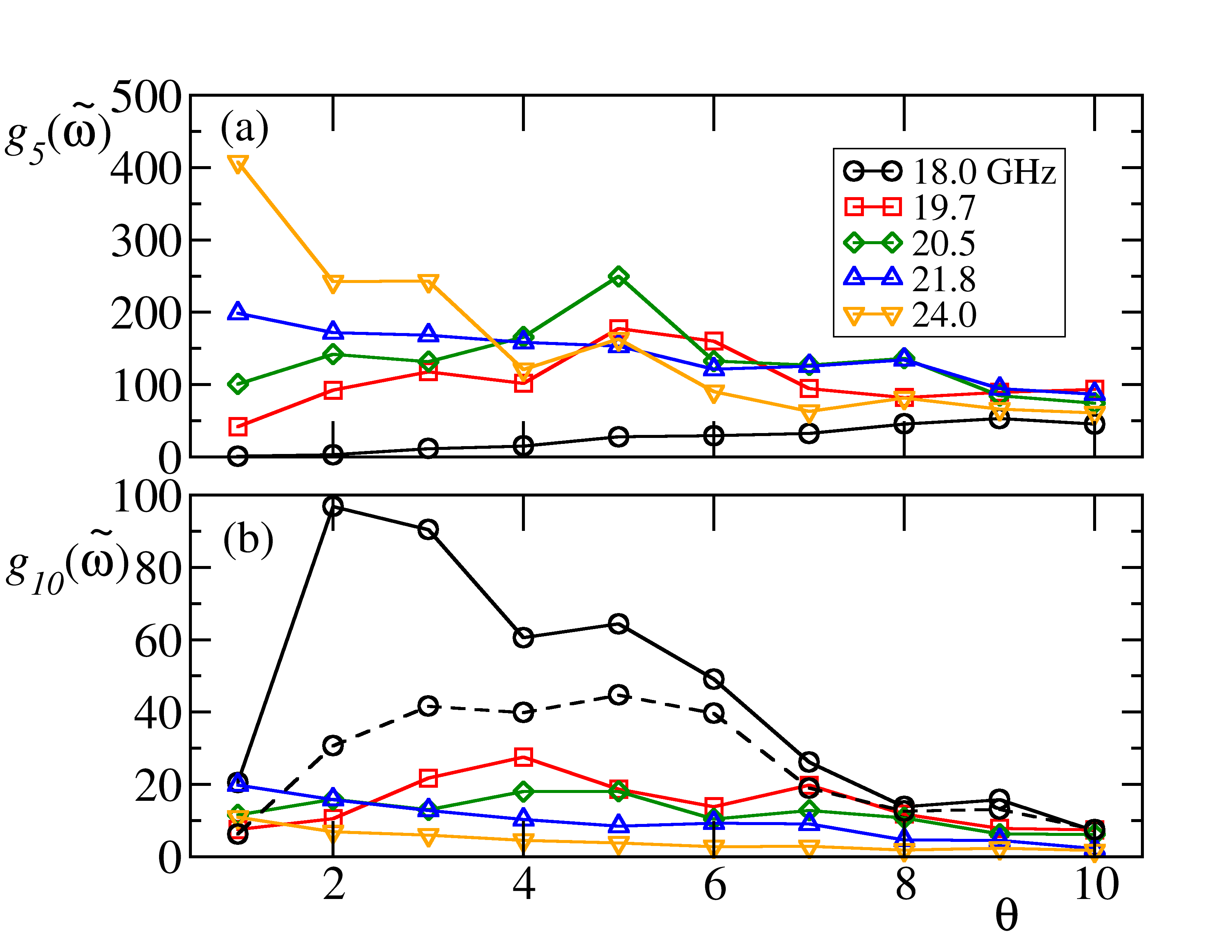}
\end{center}
\caption{Average amplitude contributions $g_n(\tilde\omega)$ on site $n=5$ (\textbf{a}) and site $n=10$ (\textbf{b}) for 
$\tilde\omega=\omega_1, \omega_2,\omega_3,\omega_4, \omega_5$ computed from data of Figure~\ref{fig:figure6p}. Solid curves
are obtained with $\delta\omega=0.25$~GHz, while the black dashed curve refers to $\delta\omega=2$ GHz.
}
\label{fig:figure7p}
\end{figure}

A more quantitative analysis can be performed by computing the contribution of a certain frequency $\tilde\omega$ to the 
wavelet signal in Figure~\ref{fig:figure6p}. Specifically, we computed the parameter 
$g_n(\tilde\omega)$ which is the average of $ \langle|G_n(\omega,t)|^2\rangle$ over the whole observation time and
over a small frequency interval centered on $\tilde\omega$ and length equal to $\delta\omega \ll \tilde\omega$.
Figure~\ref{fig:figure7p} shows the behavior of $g_n(\tilde\omega)$ for the set of characteristic frequencies 
$\omega_1,\cdots,\omega_5$ for sites $n=5$ and $n=10$ and $\delta\omega=0.25$ GHz (solid lines).
A large nonmonotonic contribution of $\tilde\omega = \omega_1$ is found on site $n=10$ which is absent for site $n=5$.
For $\delta\omega=0.25$ GHz (black solid line), the maximum amplitude is found for $\theta=2$, which is slightly smaller
than the noise amplitude that maximizes the efficiency in Figure~\ref{fig:figure5p}. 
By enlarging the averaging frequency window to $\delta\omega=2$ GHz (black dashed line), we observe a better agreement with
the efficiency curve. Therefore, we can conclude that the relevant contribution to the increase of transport efficiency of the chain
is related to the excitation of a relatively narrow set of frequencies in the interval $\omega = 18 \pm 1$ GHz on the output disk.

\section{Comparison with Coupled-Oscillators Model}
\label{sec:dnls}

For small precession angles, the dynamics of the array can be effectively described by a simplified model. Considering again the complex variables $\psi_n$ defined in Equation (\ref{eq:psi}), 
and performing a suitable expansion 
retaining only terms up to the order $\alpha$ and $p_n\equiv|\psi_n|^2$, 
the LLG equation for the chain can be approximated as an ensemble of coupled nonlinear oscillators which, in absence of any noise source, is of the form:  \cite{slavin09,borlenghi15a}
\be
\label{eq:dnls}
i\dot{\psi}_n = [-\omega_n(p_n)-i\Gamma_n(p_n)]\psi_n 
- \sum_{n^\prime}C_{nn^\prime}\psi_{n^\prime} \,. 
\ee
The first two terms on the right hand side of Equation~(\ref{eq:dnls}) are, respectively, the nonlinear frequencies $\omega_n(p_n)=\gamma |\bm{H}_{\rm{eff}}\cdot{\bm{z}}|$ and damping rates 
$\Gamma_n(p_n)$.
In our case, nonlinear effects are due to the dipolar (demagnetizing) field in each disk, and they are taken into account by expanding into powers of $p_n$ the frequencies and damping rates,
respectively, as $\omega_n(p_n)\approx\omega_n^0+\nu p_n$ and 
$\Gamma_n(p_n)=\alpha\omega_n(p_n)$ \cite{slavin09,borlenghi15b}.
Finally, the term $C_{nn^\prime}$ is the interlayer (complex) 
coupling due to magneto-dipolar interaction. 
In order to keep the model as simple as possible, 
in the following, we will consider the simple case of a uniform nearest-neighbor interaction which amounts to retain only terms 
containing $\psi_{n\pm 1}$ in Equation~(\ref{eq:dnls}) and set $C_{nn^\prime}\equiv J(1+\alpha)\delta_{n,n\pm1}$. 
Altogether, based on Equation~(\ref{eq:dnls}) and the above simplifying assumption, 
we consider the following model 
\begin{equation}
i\dot{\psi}_n = (1+i\alpha)\left[-\nu |\psi_n|^2\psi_n -\omega_n^0\psi_n
 -J(\psi_{n-1}+\psi_{n+1}) \right] +\theta^\prime\eta_n(t) \psi_n \quad,
 \label{eq:dnls2}
\end{equation}
where we also added a pure-dephasing (multiplicative) noise controlled by the parameter $\theta^\prime$ 
that acts independently on each oscillator, as well as where $\eta_n(t)$ is a Gaussian noise with zero mean and unit variance satisfying 
$\langle \eta_n(t)\eta_m(t')\rangle=\delta_{mn}\,\delta(t-t')$ (noise can be 
interpreted in the usual Stratonovich sense).
In the $\theta^\prime=0$ limit, the model is the (dissipative) Discrete Nonlinear Schr\"odinger
equation (with site-dependent frequencies). The form of dissipation in 
Equation~(\ref{eq:dnls2}) ensures that the system reaches thermal equilibrium 
when put in contact with a Langevin thermal bath at temperature $T$ \cite{iubini13,Iubini2017}.

To account for the mirror symmetry exhibited by the micromagnetic system, we consider a set of linear frequencies $\omega_n^0$ such that $\omega_n^0=\omega_{N-n}^0$. Upon normalizing the frequency $\omega_1^0$ of the first oscillator to 1, we choose non-dimensional parameters $J=0.1$,
$\alpha=0.008$ and $\nu=1$. These parameters provide a quite reasonable coarse-grained description of the out-of-equilibrium
dynamics of macrospin system, as shown in Reference \cite{borlenghi15b}.

\begin{figure}[H]
\begin{center}
\includegraphics[width=10cm]{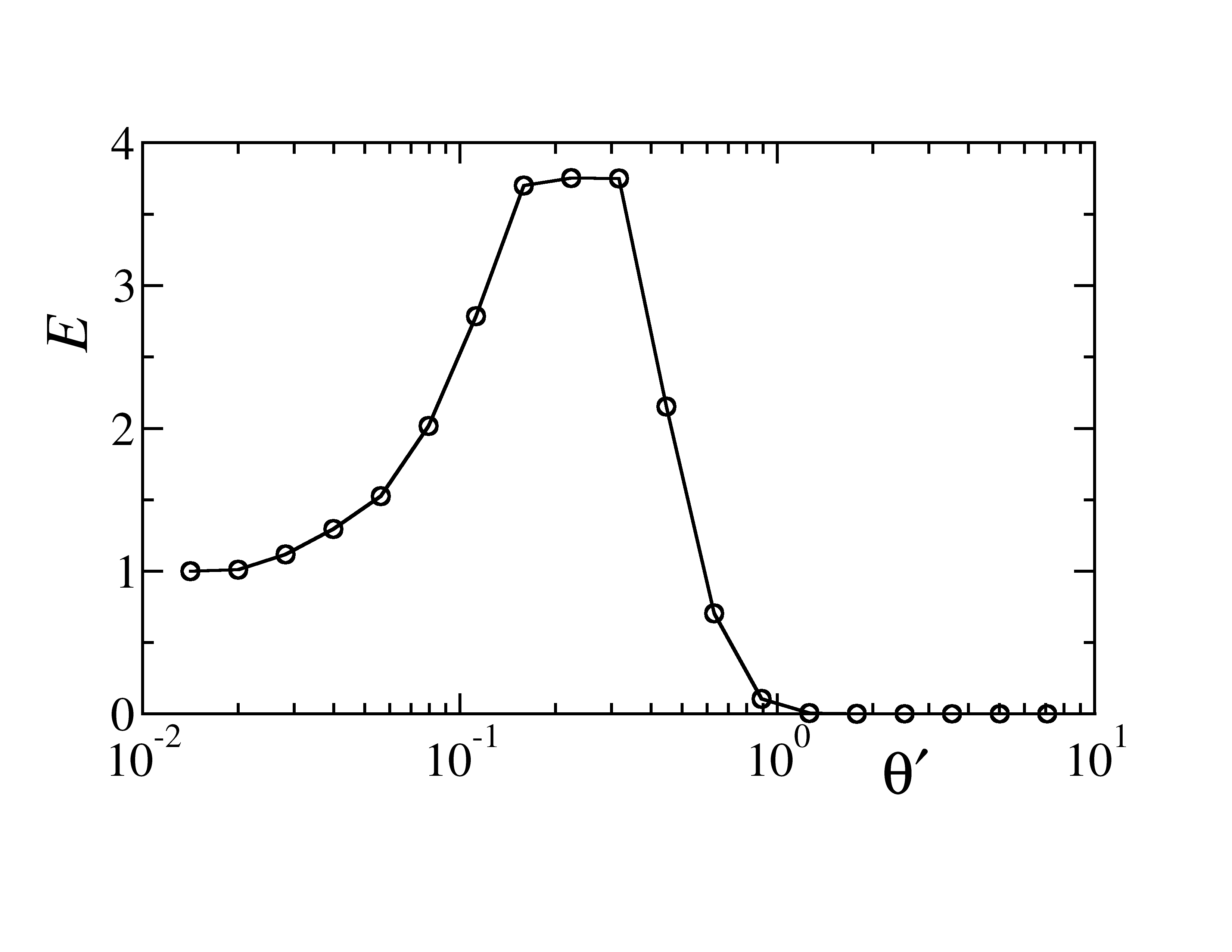}
\end{center}
\caption{
Transport efficiency versus dephasing noise strength of a chain of $N=10$ damped discrete nonlinear Schr\"odinger equation (DNLS) oscillators evolving according to Equation (\ref{eq:dnls2}). Simulations refer to $J=0.1$, $\alpha=0.008$, $\nu=1$ and linear frequencies 
$(\omega_1^0,\omega_2^0,\omega_3^0,\omega_4^0,\omega_5^0)= (1, 1.09, 1.15, 1.21, 1.33)$. For each value of $\theta^\prime$, data are averaged over a set of 100 independent realizations of the dynamics. 
}
\label{fig:figure8p}
\end{figure}

The DNLS transport efficiency $E$ is shown in Figure \ref{fig:figure8p}, as
a function of the noise strength $\theta^\prime$.
Simulations are performed by initialising the DNLS chain in a state where
the first site is excited with norm $|\psi_1|^2=1$ and the rest of the chain is empty. The system is then evolved
in the presence of dephasing noise until the total norm of the system goes below the threshold value $10^{-5}$. 
It was verified that this threshold provides a reasonable approximation of the integral in Equation (\ref{eq:efficiency}). 
A clear maximum is visible 
as in the micromagnetic case, confirming that, in spite of all the 
crude approximations, the DNLS model captures the basic features of the original setup. 
Within this simplified dynamical picture, the onset of optimal transport efficiency in the disk chain can be rationalized as follows. 
For vanishing $\theta^\prime$, the spatial heterogeneity of the local frequencies $\omega_n^0$ produces 
spatially localized normal modes~\cite{Lepri03}, which are almost entirely damped by the internal dissipation of the chain, with no
net transport to the output region. On the other hand, in the limit $\theta^\prime \rightarrow\infty$, the norm current
$J \langle Im(\psi_n^*\psi_{n+1}) \rangle$~\cite{borlenghi15b} is effectively suppressed by the effect of the local dephasing noise 
irrespective of the disks dynamics.
Optimal transport is, therefore, attained for intermediate values of the noise amplitude which provide a partial suppression of the 
localization mechanism~\cite{rayanov13,iubini15}.

\section{Conclusions}

In conclusion, we demonstrated a new phenomenon, dephasing-assisted spin transport, which shows how environmental noise can contribute 
to increasing the transfer of excitations in out-of-equilibrium spin systems. 

We argued that the 
effect is due to the subtle effect of noise on the dynamics of spin-wave modes, 
as evidenced by the spectral analysis via the wavelet transform. 
The optimal transmission occurs at noise level large enough to excite more 
extended modes without degrading their overall coherence. 
This phenomenology
is analogous to environment-assisted transport enhancement observed in 
open quantum systems~\cite{rebentrost09,roos19} and originates from the emergence of an effective macroscopic Spin Wave amplitude which
evolves according to a dissipative DNLS equation. 
On the one hand, this points towards
new energy-efficient spintronics and magnonics devices. On the other hand, the generality of the DNLS model strongly suggests that
dephasing-assisted transport is a general phenomenon that can be observed in a large class of classical
oscillating systems. 

{Although our study has focused entirely on a transient nonequilibrium dynamics, we expect that dephasing-assisted transport can show up also in
stationary conditions, i.e., when the disk chain steadily exchanges energy and magnetization between two external reservoirs~\cite{borlenghi15b}. 
In the latter setup, which appears to be more accessible to experimental tests, stationary currents should depend non-monotonously on the dephasing 
noise strength. 
}

Admittedly, our results are mostly phenomenological so far .
A more detailed theoretical study is needed and will be the subject of future work.

\authorcontributions{S.B. and S.I. performed the simulations. All Authors contributed to the research work and to writing the paper.
}

\acknowledgments{{S.L. acknowledges partial support from project MIUR-PRIN2017} {Coarse-grained description for non-equilibrium systems and transport phenomena (CO-NEST)} n. 201798CZL. S.I. acknowledges support from Progetto di Ricerca Dipartimentale BIRD173122/17 of the University of Padova.
A.D. and S.B. acknowledge financial support from Swedish e-science Research Centre (SeRC), Vetenskapsradet, the Swedish Energy Agency, and the Knut and Alice Wallenberg Foundation. Some of the computations were performed on resources provided by the Swedish National Infrastructure for Computing (SNIC) at the National Supercomputer Center (NSC), Linkoping University, the PDC Centre for High Performance Computing (PDC-HPC), KTH, and the High Performance Computing Center North (HPC2N), Umea University.
} 
\conflictsofinterest{The authors declare no conflict of interest.} 

\abbreviations{The following abbreviations are used in this manuscript:\\

\noindent 
\begin{tabular}{@{}ll}
SW & Spin Wave\\
DNLS & Discrete Nonlinear Schr\"odinger\\
LLG & Landau Lifshitz Gilbert\\

\end{tabular}}

\end{document}